# Arbitrary multi-site two-photon excitation in four dimensions

Vincent Ricardo Daria[1,2,a)], Christian Stricker[2], Richard Bowman[3],
Stephen Redman[2] and Hans-A. Bachor[1]

[1]*ARC Centre for Quantum-Atom Optics, Australian National University, 0200 ACT, AUSTRALIA*

[2]*The John Curtin School of Medical Research, Australian National University, 0200 ACT, AUSTRALIA*

[3]*Department of Physics and Astronomy, SUPA, University of Glasgow, Glasgow G128QQ, Scotland, UK*

We demonstrate dynamic and arbitrary multi-site two-photon excitation in three-dimensions using the holographic projection method. Rapid response (fourth dimension) is achieved through high-speed non-iterative calculation of the hologram using a video graphics accelerator board. We verify that the projected asymmetric spot configurations have sufficient spatiotemporal photon density for localized two-photon excitation. This system is a significant advance and can be applied to time-resolved photolysis of caged compounds in biological cells and complex neuronal networks, non-linear micro-fabrication and volume holographic optical storage.

PACS: **87.64.-t** Spectroscopic and microscopic techniques in biophysics and medical physics;
**42.40.Jv** Computer-generated holograms; **42.30.-d** Imaging and optical processing;
**87.64.mn** Multiphoton;

a) Electronic mail: vincent.daria@anu.edu.au

Advanced microscopy and diagnostic techniques require full three-dimensional (3D) access to the sample with fast dynamic control. Developments in array illumination schemes have facilitated simultaneous control and manipulation of multiple micro-particles[1-3]. Two-dimensional (2D) patterned illumination techniques have also been used for high-speed 3D image rendering of a sample. Techniques based on programmable array microscopy[4], structured illumination[5], rotating Nipkow disks[6-8] are methods using 2D excitation patterns, confocal detection schemes and post-processing algorithms to render fluorescent samples in 3D.

However, some experiments require an intrinsic 3D localized excitation response prior to detection. This can be achieved via non-linear two-photon excitation (2PE)[9]. Such experiments (e.g. photolysis of biomolecules) also require multiple excitation sites at arbitrary positions in 3D. The challenge is therefore to design a system that provides for arbitrary multi-site 2PE and meets the response time of biological tissues and cells, typically ~1 ms.

In this letter, we have expanded and combined the techniques of high-speed holographic beam control and non-linear multi-photon absorption to demonstrate a fully controllable multi-site 2PE in four dimensions. Our system significantly advances the implementation of 3D non-linear microscopy[7-8], volume holographic storage[10], microfabrication[11] and nano-surgery[12]. In neuroscience, our system allows for realistic simulation for summating simultaneous synaptic signals from multiple sites within complex dendritic trees of a neuron. Recently, multi-site single-photon (1P) photolysis in neuronal networks has been demonstrated[13]. However, 1P absorption is a linear process and suffers from poor 3D localized excitation.

Highly localized 2PE within a 3D diffraction-limited focal spot is achieved at high spatiotemporal photon densities provided by a focused femtosecond pulsed laser. Time-multiplexed multi-site 2PE has been shown to work with high-speed scanning via acousto-optic modulators (AOM)[14-15]. However, dispersion through an AOM crystal degrades the spatiotemporal quality of the femtosecond pulsed laser, especially when four AOM's are used to project the focal spot in arbitrary 3D positions[16]. Multi-site 2PE along a plane has also been demonstrated with the holographic method that uses an iterative-adaptive optimization algorithm to produce photon efficient spot arrays[17]. They also incorporate a general lens function to shift the whole plane containing the spot array along the optical axis. On the contrary, we show a fast and arbitrary multi-site 2PE in all three dimensions. Each excitation site has a localized response in 3D. We derive the hologram using a non-iterative method and achieve optimal speeds by taking advantage of the parallel computing capability of a graphics accelerator board.

A phase hologram for 2PE at $N$ arbitrary sites can be calculated with no iterative optimization procedure by the superposition of $N$ fields[1,2]. Each field is a combination of prism and lens phase functions that describe the 3D position of a focal spot. The localized 2PE profile is derived by the square of the intensity distribution of the excitation focal spot. Aberrations disturb the excitation spots when they are arbitrarily positioned around the sample. To visualize these aberrations, we use the non-optimized holograms as input and calculate the 3D intensity profile using the Fresnel diffraction integral[18]. Fig. 1(a) shows the intensity distribution along the $xz$-plane with two spots ($N$=2) symmetrically positioned (0, 0, $f \pm 5/\xi$) with respect to the nominal focus ($f$) of the lens. The normalized transverse ($x,y$) and axial ($z$) coordinates are related to spatial coordinates by $\eta = NA\lambda^{-1}$ and $\xi = NA^2\lambda^{-1}$, respectively, where $NA$ is the numerical aperture of the objective lens and $\lambda$ is the wavelength of the laser. For $N$=2, the maximum normalized intensity of each spot is 0.405. The axial intensity distribution of the spots is characterized by aberrations with uneven side-lobes along the $z$-axis. These aberrations increase as the spots are positioned away from the focal plane of the lens. Fig. 1(b) shows $N$=4 spots positioned at (±2.25/$\eta$, 0, 5.25/$\xi$) at the top layer and (±2/$\eta$, 0, -5/$\xi$) at the bottom layer. For $N$=4, the maximum normalized intensity of each spot reduces to 0.2. No significant aberrations were observed when the spots are moved along the transverse direction. However, for highly symmetric spot configurations produced from non-optimized holograms, interference from higher diffraction orders affects the uniformity of the maximum spot intensities[19]. Upon implementation, encoding such holograms on a spatial light modulator (SLM) will introduce further losses[20]. Nonetheless, these losses and differences in spot intensities can be accounted for via optical theory.

More uniform spot intensities can be achieved via asymmetric and random spot configurations. Fig. 1(c) shows the maximum intensities (squares) for randomly arranged spots, which is inversely proportional with $N$ and follows a fit given by, $I_{max} \approx 0.84 N^{-1}$. Fig. 1(c) also shows the total intensity (triangles), which is derived by taking the sum of the maximum spot intensities for every configuration. It is interesting to note that the total intensity is around ~0.8 and slowly decreasing over a range of $N$. This indicates that losses due to higher diffraction orders are maintained at ~0.2 even as $N$ is increased. The plot for the maximum two-photon fluorescence intensity (diamonds) is also shown for reference. Fluctuations in fluorescence intensity are more evident due to the non-linear 2PE response.

We calculate the holograms using an OpenGL shader language program running on a graphics accelerator board (nVidia GeForce 9800 GT)[21]. Fig. 1(d) shows the update rate for hologram array sizes (600x600) and (1000x1000). For $N$<100, calculation of a 600x600 hologram can accommodate update rates of up to 300 frames per second (fps). For comparison, we show a significantly slower hologram calculation using a program developed in LabView$^{TM}$ v8.5 (National Instruments) and running on a desktop computer (Intel Core 2 Duo 3 GHz).

Fig. 2 describes the geometry of our optical setup, which makes use of a linearly polarized TEM$_{00}$ laser beam from a Ti:S femtosecond pulsed laser (Coherent Inc.). Lenses L1 and L2 expand the beam to illuminate the 16×12 mm$^2$ area of the phase-only SLM (Hamamatsu Photonics). The overall efficiency of the SLM is 92% with a fill factor of 95%. The update rate is currently limited by the SLM's response time of 30 ms. Electronic addressing of each pixel in the SLM is achieved via the video output of the graphics accelerator board.

The SLM operates in reflection mode, and separating the incident and the phase-encoded reflected beam with some degree of tilt optimizes the throughput. We minimize the tilt by re-using lens L2 to serve as the Fourier transform lens. The spots are formed at the focal region of L2 and relayed to microscopic scale using L3 and the objective lens. A dichroic mirror reflects the phase encoded near-infrared beam to the microscope while allowing visible light to pass through for bright field and fluorescence imaging. A standard microscope configuration with an appropriate filter is used to view the fluorescence along the *xz*-plane.

To test arbitrary multi-site 2PE along the focal plane, we prepared 2 μm diameter fluorescence latex microspheres (Molecular Probes, Inc) diluted with water. The microspheres emit green fluorescence following 2PE from a focused near-infrared fs-pulse laser. Fig. 3 shows multiple excitation site configurations. The excitation spots are arbitrarily positioned around the optical axis, which is at the center of the micrograph. The average power of each excitation spot, including system losses, is approximately 2.9, 2.0, 1.5 and 1.2 mW for *N*=15, 20, 25 and 30, respectively. System loss of around 10% is associated with light diffracted due to the physical constraints of the SLM[20]. Tolerable variations in fluorescence intensity are due to slight off-center illumination on the beads as well as slight differences in the maximum spot intensity as described earlier.

Next, we demonstrate multi-site 2PE along the optical axis. Fig. 4(a) shows a photograph of the setup where we used a 63x (*NA*=0.9) objective lens for projecting the excitation spots and a 10x lens for viewing the fluorescence from the side. We prepared a dye solution from carboxyfluorescein dissolved in water and sodium hydroxide. Fig. 4(b) shows two localized 2PE sites positioned symmetrically at $x = \pm 10$ μm. Also shown are two (Fig. 4(c)) and four (Fig. 4(d)) 2PE sites projected along the optical axis. With *NA*=0.9, the maximum displacement along the *z*-axis is around ~100 μm. Since there is an overlap between the absorption ($\lambda_{abs}$=492) and emission ($\lambda_{em}$=517 nm) spectra of carboxyfluorescein, the emitted fluorescence from the 2PE process induces 1P fluorescence at neighboring regions around the excitation spot. Hence, the fluorescence detected from the side is smeared and does not reflect the true 3D localized excitation provided by the 2PE process. Nonetheless, such demonstration establishes our capacity to provide arbitrary and localized 2PE along the optical axis.

When applied to photolysis of caged neurotransmitters, irregular and asymmetric spot distributions match the arbitrary dendritic trees of neurons. Hence, high variations in spot intensities associated with symmetric spot arrays will not pose a major problem. It is also important to note that the programmability of the SLM should allow for appropriate wavefront corrections to rectify aberrations at *z*-positions further from the focal plane. Such corrections will be worked out in future studies.

We have demonstrated multi-site and highly localized 2PE in four dimensions. Non-iterative calculations carried out with the graphics accelerator board generate holograms on-the-fly with the update rate currently limited by the response time of the SLM. Multi-site fluorescence emission indicate that the spots have sufficient spatiotemporal photon density for 2PE. For quantitative measurements, standard optical theory can account for slight intensity differences introduced by non-optimal holograms. Immediate applications of our system point to neuroscience, where photo-stimulation of neurotransmitters at arbitrary sites provides fundamental answers to the mechanisms of information processing in the human brain.

This project has been funded by the Australian National University – Office of the Vice-chancellor and the major equipment grant (08MEC06). The work of HAB was supported by the Australian Research Council Centre of Excellence scheme. We thank Miles Padgett of the University of Glasgow for sharing the software for high-speed calculation of holograms.

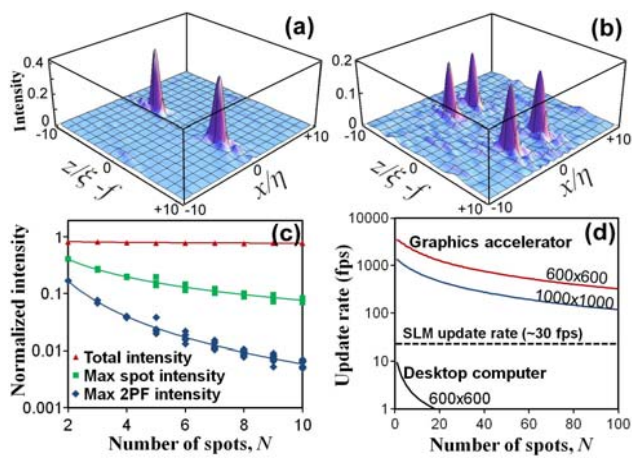

Figure 1. Intensity distribution for (a) two and (b) four spots along the *xz*-plane. (c) Plot as a function of number of spots, *N*, of the total intensity (triangles), maximum spot intensity (squares) and two-photon fluorescence (2PF) (diamonds), while (d) is the hologram update rate.

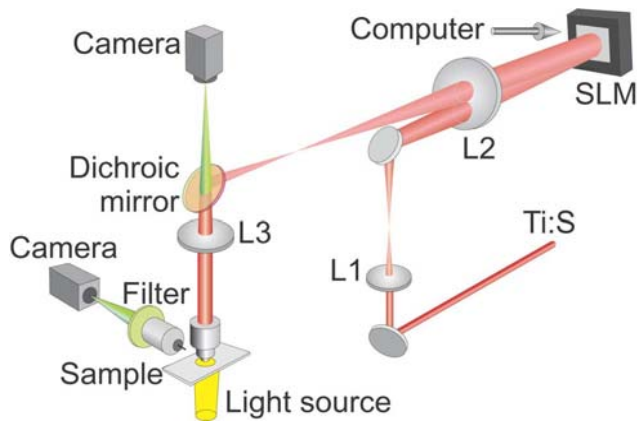

Figure 2. Optical setup for multi-site two-photon excitation.

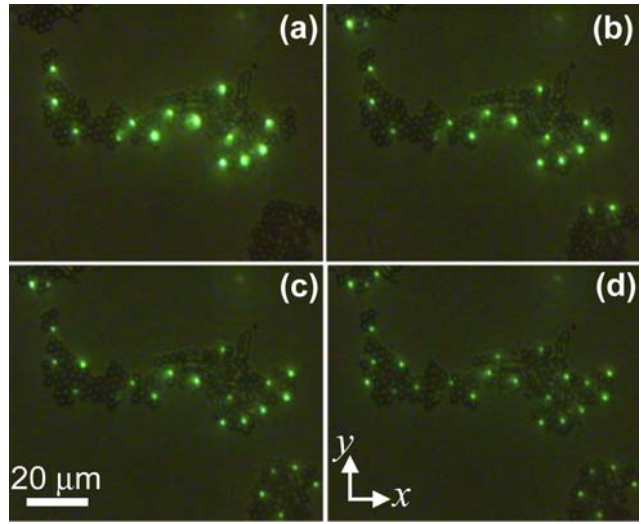

Figure 3. Arbitrary multi-site 2PE on fluorescent latex microbeads along the focal plane for: (a) *N*=15; (b) *N*=20; (c) *N*=25; (d) *N*=30.

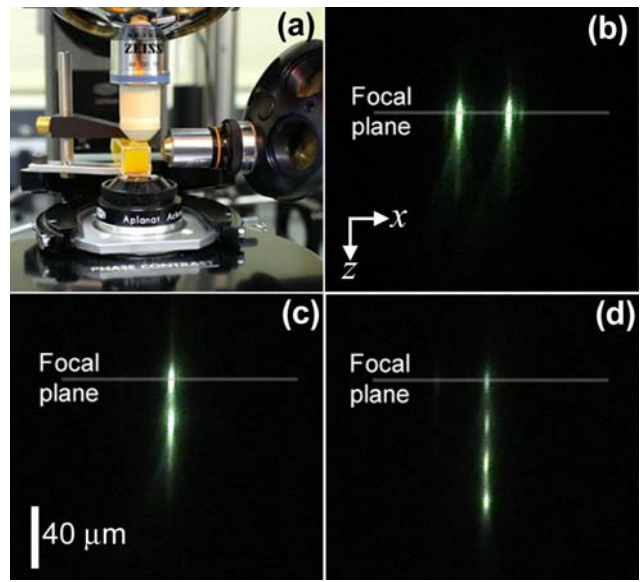

Figure 4. Arbitrary multi-site 2PE along the optical axis. (a) Setup to view the *xz* - fluorescence distribution. Localized fluorescence with (b) two sites within the focal plane; while (c) two and (d) four sites along the optical axis.